\documentclass[usegraphicx,letterpaper,usenatbib]{mn2e}  
\usepackage{amsmath,amssymb,latexsym,multirow,longtable,threeparttablex}
\usepackage[utf8]{inputenc}
\usepackage{graphicx}
\usepackage{hyperref}
\usepackage{aas_macros}

\usepackage[totalwidth=480pt,totalheight=680pt,layoutvoffset=0.5cm]{geometry}

\usepackage[usenames]{color}

\newcommand{\tempo}{{\sc tempo2}}
\newcommand{\bayesfit}{{\sc bayesfit}}

\begin{document}
\title{Bayesian inference for pulsar timing models}
\author[S.\ J. Vigeland and M.\ Vallisneri]{Sarah J.\ Vigeland$^1$ and Michele Vallisneri$^1$ \\
$^1$Jet Propulsion Laboratory, California Institute of Technology, Pasadena, CA 91109, USA}
% revtex - \author{Sarah J.\ Vigeland}
% revtex - \author{Michele Vallisneri}
% revtex - \affiliation{Jet Propulsion Laboratory, California Institute of Technology, Pasadena, CA 91109, USA}

\maketitle
\date{\today}
\begin{abstract}
The extremely regular, periodic radio emission from millisecond pulsars makes them useful tools for studying neutron star astrophysics, general relativity, and low-frequency gravitational waves. These studies require that the observed pulse times of arrival be fit to complex timing models that describe numerous effects such as the astrometry of the source, the evolution of the pulsar's spin, the presence of a binary companion, and the propagation of the pulses through the interstellar medium. In this paper, we discuss the benefits of using Bayesian inference to obtain pulsar timing solutions. These benefits include the validation of linearized least-squares model fits when they are correct, and the proper characterization of parameter uncertainties when they are not; the incorporation of prior parameter information and of models of correlated noise; and the Bayesian comparison of alternative timing models. We describe our computational setup, which combines the timing models of \tempo~with the nested-sampling integrator {\sc MultiNest}. We compare the timing solutions generated using Bayesian inference and linearized least-squares for three pulsars: B1953+29, J2317+1439, and J1640+2224, which demonstrate a variety of the benefits that we posit.
\end{abstract}

% Note: according to the NANOGrav publication policy (http://wiki.nanograv.org/index.php?title=Publication_Policy_Adopted_on_August_17_2012):
% this is a Non-NANOGrav Science Publication (5.a.iii) which uses NANOGrav resources but is not on topics related to NANOGrav science;
% by 6.c, publications by NANOGrav members which use NANOGrav resources (data or tools) for non-NANOGrav science are subject to the authorship policies governing NANOGrav data and NANOGrav tools given in Section 6.d but are not governed by other authorship rules in the NANOGrav publication policy;
% by 6.d.iii, publications that use small amounts of published NANOGrav data for illustrative purposes, but which do not present new GW limits/detections, need not include data-takers among the authors.
% Therefore

% revtex - \pacs{95.75.Pq, 97.60.Gb, 97.80.Af}
% revtex - \maketitle

\section{Introduction}
\label{sec:intro}

With spin periods as stable as a part in $10^{15}$ over years, millisecond pulsars are exceedingly regular rotators, and their periodic radio pulses can be timed with sub-$\mu$s accuracy \citep{2008ApJ...679..675V},
%\citep{2008LRR....11....8L,2012hpa..book.....L}
making pulsars great laboratories to study neutron-star astrophysics, to test general relativity (\citealt{2006Sci...314...97K,2013Sci...340..448A}; for a review, see \citealt{2003LRR.....6....5S}) 
and to serve as quasi-ideal clocks in searches for gravitational waves (GWs: \citealt{1978SvA....22...36S,1979ApJ...234.1100D}). The most sensitive GW searches proceed by correlating the unexplained \emph{residuals} in the timing series of multiple pulsars in a \emph{pulsar-timing array} (PTA: \citealt{1983ApJ...265L..39H}).
Recent years have seen a renewed interest in this effort, with the formation of three major regional programs [the EPTA \citep{2011MNRAS.414.3117V}, NANOGrav \citep{2013ApJ...762...94D}, and the PPTA \citep{2013Sci...342..334S}], which have now joined into a global collaboration to share data and expertise \citep{2010CQGra..27h4013H}.

A large part of the science enabled by pulsars unfolds through a rather pure instance of \emph{model fitting}: pulsar astronomers build complex, many-parameter mathematical models of pulsar-spin evolution and radio-pulse propagation out of the pulsar system (which is often a close binary), through interstellar space, across the Solar system, and into our large radio dishes and broadband receivers. These \emph{timing models} must include not only the basic geometry of light propagation, but also the effects of gravity, relativity, binary companions, and more. They must also predict the times of arrival of every single pulse (out of hundreds of millions across years of observation, although only a subset of pulses are actually timed) within a fraction of a period \citep*{2013CQGra..30v4001L,2006MNRAS.372.1549E}. It is no wonder then that such a painstaking, all-encompassing effort should yield physical measurements of great precision, although one may wonder that Nature would be so kind to gift such excellent toys to astronomers!

In practice, model fitting must be preceded by the proper calibration of the data and by the mitigation of systematic effects, both of which are major areas of research (see \citealt{2013CQGra..30v4001L} for some references). One interesting twist is that timing models must be \emph{bootstrapped}, since pulses from millisecond pulsars are not detectable individually, but must be ``folded'' in batches of hundreds or thousands, requiring a basic (but already accurate) model to align the folding. Another twist is that, for arguably the most promising GW source for PTAs, the stochastic background from the ensemble of massive--black-hole binaries at the centers of galaxies \citep{2013MNRAS.433L...1S}, the signal of interest is the \emph{noise-like} residual that is left after subtracting the deterministic timing model, which can only be characterized statistically. Luckily the GW background is distinguished from other sources of noise by appearing in correlated fashion for all the pulsars in the array \citep{1983ApJ...265L..39H}.

After the pulsar datasets are distilled (by way of rather sophisticated processing) into sets of \emph{times of arrival}, the estimation of pulsar parameters is routinely performed by solving for the \emph{maximum-likelihood} timing model using \emph{linear least squares} \citep[ch.\ 15]{1992nrca.book.....P}, as implemented in the widely adopted pulsar-timing software packages \textsc{tempo} \citep{tempoweb} and \textsc{tempo2} 
%\citep{tempo2web,2006MNRAS.369..655H,2006MNRAS.372.1549E,2009MNRAS.394.1945H}
(\citealp*{tempo2web,2006MNRAS.369..655H}; \citealp{2006MNRAS.372.1549E,2009MNRAS.394.1945H}). These packages are impressive accomplishments in the complexity and precision of their timing models, but they offer only rather straightforward fitting options.

The main limitation of linear--least-squares maximum-likelihood estimation is that it is a \emph{local} technique that relies on linearizing the model of interest with respect to a vector of parameter corrections.
This does not keep the technique, in most practical cases, from finding global maxima by chaining a sequence of local corrections.
It does however mean that the linear approximation must be accurate at least across the parameter ranges returned as uncertainties by the least-squares process itself---otherwise, the reported uncertainties are certain to be unreliable. Thus, the risk is that the impressive accuracy claimed in some pulsar-timing ephemerides could be overstated, especially for poorly constrained parameters.

The other limitation of the standard approach does not concern the mathematical technique itself, but rather its physical boundary conditions. Least-squares timing-model fits are usually set up by assuming that the only source of error in the times of arrival is \emph{radiometer noise}, which can be estimated directly from the raw data (more in Sec.\ \ref{sec:timing}), and which is taken to be Gaussian and uncorrelated. However, for certain pulsars these errors appear to be too large, or too small (i.e., within the logic of least-squares estimation, they lead to very small or very large $\chi^2$ values), thus skewing uncertainty estimates. Other pulsars show hints of residual correlated noise with ``red'' spectra dominated by low-frequency components \citep{2010arXiv1010.3785C,2010ApJ...725.1607S,2011MNRAS.418..561C,2013MNRAS.428.1147V}, which is not taken into account in the standard estimation process. Even the very presence of GWs could conceivably bias parameter estimation.\footnote{Although this would be a welcome annoyance, because it would imply rather large GW amplitudes and imminent detections.}

In this paper we argue that all these concerns can be addressed by a fully nonlinear Bayesian-inference approach \citep{2010blda.book.....G} that derives \emph{distributions} (rather than \emph{point estimates}) for the pulsar-timing parameters by sampling freely across parameter space, and computing the full timing model at each point. Furthermore, we argue that this approach offers additional scientific opportunities (listed below), and that it is both practical and computationally feasible with current hardware resources and software components. As an example, in Sec.\ \ref{sec:results} we present Bayesian analyses of three NANOGrav pulsars \citep{2013ApJ...762...94D}, which we performed by using \textsc{tempo2} as a library to form timing residuals for arbitrary pulsar parameter sets, and by calling it from a popular stochastic sampler [the nested-sampling \citep{2004AIPC..735..395S} package MultiNest \citep*{2009MNRAS.398.1601F}].

Our experience leads us to advocate for a broad use of full Bayesian inference in timing-model studies, because of the following benefits:
\begin{itemize}
\item \emph{Uncertainties.} Bayesian inference offers reliable estimates of parameter uncertainties (as the moments or quantiles of marginalized posteriors) and parameter correlations (as the joint marginalized posteriors). These estimates fully take into account the nonlinear functional dependence of the model on the parameters. The least-squares formalism can reproduce these estimates only in the infinite signal-to-noise limit where all parameters have normal error distributions around their maximum-likelihood values.
\item \emph{Priors.} Bayesian inference allows the statistically principled incorporation of prior knowledge about some of the parameters. For instance,
% Madison and colleagues
\citet*{2013ApJ...777..104M} argue that pulsar positions and proper motions determined from VLBI astrometry can substantially improve timing-model determinations. In our worked-out examples of Sec.\ \ref{sec:results}, we use estimates of distance from dispersion measurements as priors for the parallax, a poorly determined parameter that can even take unphysical negative values in least-squares estimation.
\item \emph{Noise.} Bayesian inference lets us solve simultaneously for the timing-model parameters and for the noise parameters that describe the errors in the timing data, obtaining joint probability distributions for both. As discussed by
% van Haasteren and Levin
\citet{2013MNRAS.428.1147V}, we may model the errors as containing a white, uncorrelated noise component (perhaps with extra parameters that modulate the empirical estimates of radiometer noise) but also any number of red, correlated Gaussian processes \citep{Rasmussen2005}.
\item \emph{Model comparison.} Bayesian inference lets us compare and choose between alternative timing+noise models with different parameter ranges or components, by evaluating the relative model \emph{evidence} \citep{2010blda.book.....G}. The comparison may be, for instance, between a white-noise and white+red-noise model, two different descriptions of binary dynamics, or even the dynamics of general relativity and that of a modified theory of gravity
% (as advocated by Del Pozzo and colleagues for GW measurements \citep{2011PhRvD..83h2002D}).
[as advocated by \citet*{2011PhRvD..83h2002D} for GW measurements].
\end{itemize}
We note one last benefit: we can use nonlinear Bayesian inference to \emph{validate} the results of least-squares, maximum-likelihood estimation, which by itself does not offer the means for such a validation, but which will undoubtedly remain the workhorse in this field. Such validation is necessary also for the noise-estimation and GW-search methods  that marginalize the posterior over the timing-model parameters by assuming that the residuals can be accurately represented as linearized functions \citep{2011MNRAS.414.3117V}; we shall say more about these methods in Sec.\ \ref{sec:timing}.

We are not the first to suggest that Bayesian methods can be applied to pulsar timing.
% Messenger and colleagues \citep{2011CQGra..28e5001M}
\citet{2011CQGra..28e5001M}
propose two Bayesian methods to generate times of arrival from raw and folded radio data---that is, to perform the data-analysis step prior to timing-model fitting. They also discuss the possibility of an end-to-end Bayesian analysis. Independently from and simultaneously with our work,
% Lentati and colleagues \citep{temponest}
\citet{temponest} developed a software package, based on the very same two components that we use, to perform a joint Bayesian analysis of the nonlinear pulsar-timing model and of any stochastic components. They emphasize that the concurrent modeling of correlated noise (in addition to uncorrelated time-of-arrival errors) can change the estimates and uncertainties of timing-model parameters very significantly; working with simulated and real pulsar data, they show examples of non-normal parameter posteriors in observations with low signal-to-noise ratio, and they compare the Bayesian evidences of different noise models and different subsets of timing-model parameters. We agree with most of the theoretical framework laid out in \cite{temponest}, but we differ on the proper treatment of Bayesian model comparison, as we discuss in greater detail in Sec.~\ref{sec:conclusions}.

The rest of this paper is organized as follows. In Sec.\ \ref{sec:timing} we present an overview of pulsar--timing-model fitting of both least-squares and Bayesian flavors, and explain how the latter overcomes the shortcomings of the former. In Sec.\ \ref{sec:compute} we describe our computational setup, and in Sec.\ \ref{sec:results} we recount our three case studies of Bayesian model fitting. Last, in Sec.\ \ref{sec:conclusions} we offer our conclusions and recommendations.

\section{An overview of pulsar timing and model fitting}
\label{sec:timing}

While it is useful to think of pulsars as ideal clocks for the purpose of GW detection, there are actually several experimental steps that separate the emission of radio pulses from the analyses that can determine pulsar parameters and to search for GWs.
% Lorimer \citep{2008LRR....11....8L} and Lommen and Demorest \citep{2013CQGra..30v4001L}
\citet{2008LRR....11....8L} and \citet{2013CQGra..30v4001L}
give thorough accountings of all the steps involved, which we summarize in this section to provide context for our later discussion. In this section we also develop the basic mathematical formalism needed for standard (i.e., least-squares) and Bayesian model fitting.

\paragraph*{Times of arrival.} After the pulses are emitted, they traverse the interstellar medium and are received at the radio telescope, where they are usually observed at central frequencies $\sim 1$ GHz, with bandwidths of tens to hundreds of MHz. The radio signals are then dedispersed and \emph{folded}: that is, because the pulses are too weak to be analyzed individually, they are aligned (requiring a basic fiducial set of pulsar parameters) and added together. For many pulsars, the resulting average pulse shapes have remarkably stable \emph{profiles} across observing epochs, so they allow the accurate determination of pulse \emph{times of arrival} (TOAs). These are defined as the time\-stamps of a fiducial point on the profile, such as its maximum, referred to a (notional) individual pulse at the beginning or at the midpoint of each observation \citep{2013CQGra..30v4001L}. In effect, TOAs are determined by cross-correlating the observed profiles with an analytical or empirical ``template.''

\paragraph*{TOA errors.} The resulting TOA uncertainty is roughly the width of the profile, divided by signal-to-noise ratio (SNR) of the observation. These errors are usually idealized as Gaussian, uncorrelated noise (known as \emph{radiometer} noise), although the variation of profiles with time and across the observation bandwidth has the potential of introducing systematic components.
Furthermore, for some pulsars and observations the correlation uncertainties appear to be systematically over- or under-estimated; as a result, observers introduce \emph{ad hoc} corrections to multiply the uncertainties by a constant factor (EFAC) or to postulate additional uncorrelated noise (EQUAD) \citep{2013CQGra..30v4001L}.

\paragraph*{Timing model.} The regularity of pulsar emission becomes evident only after the \emph{topocentric} TOAs $t_i^\mathrm{obs}$ are transformed to the \emph{pulsar frame} by subtracting a number of systematic propagation delays \citep{2006MNRAS.372.1549E}. Working back from the observatory to the pulsar, these delays include:
\begin{itemize}
\item the transformation from the observatory frame to the Solar-system-barycenter (SSB) frame, which involves \emph{clock corrections} to a fiducial terrestrial time standard, the motion of the Earth, dispersion in the Earth's atmosphere and in the interplanetary medium, as well as special- and general-relativistic effects;
\item the transformation from the SSB to the emitting system, which involves the secular motion of the pulsar and dispersion in the interstellar medium;
\item for pulsars in a binary, the delays due to the massive companion, which involve Newtonian and post-Newtonian orbital dynamics, aberration, and again special- and general-relativistic effects.
\end{itemize}
These delays are complex functions of a small number of parameters that describe the geometry and kinematics of the pulsar system \citep{2006MNRAS.372.1549E}. Their computation also requires additional inputs such as a numerical Solar-system ephemeris, clock-correction tables, and troposphere data. The resulting \emph{timing model} is completed by a description of the intrinsic evolution of pulse phase, usually given as a polynomial in the pulsar-frame time $t^\mathrm{psr}$:
\begin{equation}
\phi(t) = \sum_{j \geq 1} \frac{f^{(j-1)}}{j!} (t^\mathrm{psr} - T_0)^j + \phi_0,
\end{equation}
where $f^{(k)}$ is the $k$-th derivative of spin frequency; the epoch $T_0$ is set by the observer; and $\phi_0$ is an absolute phase alignment, which is customarily measured in \emph{turns} equal to $2 \pi$ radians.
The most significant parameters in a timing model are those that describe the position of the pulsar and its intrinsic spin frequency and frequency derivative; models for pulsars in binaries also require five Keplerian parameters. Additional parameters describe the effect of dispersion in the interstellar medium, plus the more subtle effects of proper motion, parallax, dispersion in the interstellar medium, post-Keplerian dynamics, and more.

\paragraph*{Timing residuals.}
The timing \emph{residuals} are defined as time delays equivalent to the \emph{excess phase} predicted by the timing model at the pulsar-frame--transformed TOAs \citep{2006MNRAS.369..655H}:
\begin{equation}
\label{eq:res}
R_i(\theta^\alpha) = \frac{\phi(t_i^\mathrm{psr};\theta^\alpha) - N_i}{f_0};
\end{equation}
here $N_i$ is the nearest integer to each $\phi(t_i^\mathrm{psr};\theta^\alpha)$, and we indicate explicitly the dependence of the residuals on the timing-model parameters $\theta^\alpha$. Indeed, the $t_i^\mathrm{psr}$ themselves should be understood as functions $t_i^\mathrm{psr}(t_i^\mathrm{obs};\theta^\alpha)$.
The best-fitting timing model is then defined as the set of parameters $\hat{\theta}^\alpha$ that minimizes the $\chi^2$ of the residuals,
\begin{equation}
\chi^2(\theta^\alpha) = \sum_i \left(
\frac{R_i(\theta^\alpha)}{\sigma_i}
\right)^{\!2},
\end{equation}
where the $\sigma_i$ are the measurement uncertainties for the $t_i^\mathrm{obs}$ (and by extension for the $R_i$). Indeed, if we assume that radiometer noise is the only source of error, so that the $t_i^\mathrm{obs}$ are measured with uncorrelated, normally distributed errors with variance $\sigma_i^2$, then the likelihood of the data given the $\theta^\alpha$ is $\propto \exp (-\chi^2(\theta^\alpha)/2)$, and the $\chi^2$-minimizing vector $\hat{\theta}^\alpha$ is the \emph{maximum-likelihood estimator} of the pulsar parameters. The best models yield residuals as small as a milliperiod \citep{2013CQGra..30v4001L}.

\paragraph*{Least-squares minimization.}
Obtaining $\hat{\theta}^\alpha$ may seem problematic, since Eq.\ \eqref{eq:res} makes sense only if the timing model can keep track of which turn the pulsar is performing, so that the ``right'' $N_i$ is subtracted from each $\phi(t_i^\mathrm{psr};\theta^\alpha)$; such a solution is said to be ``phase connected.''
In practice, accurate models are obtained by beginning from a subset of data and from those parameters that have the strongest effect on the timing model (setting extra parameters to arbitrary values, or omitting the corresponding terms in the model), so that it is easier to maintain phase connection, and then progressively extending both sets. 
One may worry that such a procedure, which requires an experienced guiding hand, may not always converge to the globally optimal solution, but such concerns have usually been allayed by the smallness of the final residuals over long observation times.

Computationally, the approximate solution $\theta^\alpha_{(k)}$ can be improved iteratively by minimizing the linearized $\chi^2(\theta_{(k+1)}^\alpha)$,
\begin{equation}
\sum_i \left(
\frac{R_i(\theta_{(k)}^\alpha) + M_{i\beta}(\theta^\alpha_{(k)}) \delta \theta^\beta}{\sigma_i}
\right)^{\!2}, 
\end{equation}
with respect to the parameter correction $\delta \theta^\alpha = \theta_{(k+1)}^\alpha - \theta^\alpha_{(k)}$. Here $M_{i\beta}(\theta^\alpha) = \partial \phi(t_i^\mathrm{psr}) / \partial \theta^\beta |_{\theta^\alpha}$ is known as the \emph{design matrix}, and we imply a summation over the parameter index $\beta$.
Each such minimization is a weighted least-squares problem equivalent to the linear system
\begin{equation}
\label{eq:leastsquares}
M^T N^{-1} M \, \boldsymbol{\delta \theta} = -M^T N^{-1} \mathbf{R},
\end{equation}
where $N$ is the diagonal \emph{noise correlation matrix} with components $N_{ii} = \sigma_i^2$. 

\paragraph*{Convergence and statistical interpretation.}
The timing-model solution is repeatedly corrected until it converges to the maximum-likelihood estimator \citep{2006MNRAS.369..655H}, where (under high-SNR conditions that we will discuss below) the standard uncertainties of the estimated parameters are described by the covariance matrix $(M^T N^{-1} M)^{-1}$, and where the $\chi^2$ itself is drawn from the eponymous distribution with $n - m$ degrees of freedom, with $n$ the number of residuals and $m$ the number of parameters \citep[ch.\ 15]{1992nrca.book.....P}. In practice, checking whether $\chi^2 \sim n - m$ (the expected value for the distribution) is the main quantitative criterion to decide whether a solution is reasonable.
Likewise, the progressive reduction in a related quantity, the weighted rms residual $\bar{R} = (\sum_i R_i^2 \sigma_i^{-2} / \sum_i \sigma_i^{-2})^{1/2}$, is used as heuristic to decide whether to add extra fine-tuning parameters to a model.
If the $\chi^2$ is large but the model is trusted to be correct, one possible conclusion is that the estimated TOA uncertainties are too small; it is then standard practice to assume that the $\sigma_i$ are really smaller by a factor $[\chi^2 / (n - m)]^{1/2}$, which has also the effect of reducing the parameter uncertainties proportionally (this is for instance the standard behavior of the \textsc{tempo2} package).

\paragraph*{Underdetermined problems.}
We note that the linear system of Eq.\ \eqref{eq:leastsquares} is ill-defined if the columns of $M$ are not orthogonal (i.e., if there are combinations of pulsar parameters that can be changed together without any effect on the residuals). In that case, the least-squares problem can still be solved using the singular-value decomposition \citep{2006MNRAS.369..655H} if we impose the additional requirement of minimizing $||\boldsymbol{\delta \theta}||$ in addition to $\chi^2$ \citep[p.\ 676]{1992nrca.book.....P}. In an iterative-improvement scheme, this may have the effect of discouraging movement from possibly arbitrary initial values of spurious or weakly acting parameters.

\paragraph*{Bayesian inference: formulation.}
The maximum-likelihood and least-squares setup can be extended rather easily to the case of correlated TOA errors, as long as these are modeled as one or more \emph{Gaussian noise processes} \citep{Rasmussen2005}. Loosely speaking, Gaussian processes are the extension of the notion of random Gaussian variables to functions, which are then completely characterized by a covariance function $C(t_1,t_2)$. We would then write the $\chi^2$ as $\sum_{i,j} R_i (C_{ij})^{-1} R_j$, where $C_{ij} = C(t^\mathrm{obs}_i,t^\mathrm{obs}_j)$ is the total noise covariance matrix, and replace $N$ with $C$ in the least-squares linear system of Eq.\ \eqref{eq:leastsquares}.
The problem with doing so is that, while we obtain an estimate of radiometer noise from the profile-correlation process, we can seldom provide an exact specification for the other noise components. We may then adopt an extended maximum-likelihood approach that maximizes
\begin{multline}
\label{eq:likelihood}
p(t^\mathrm{obs}_i|\theta^\alpha,\eta^A) = \frac{1}{\sqrt{(2 \pi)^n |C(\eta^A)|}} \\
\times \exp \Bigl\{-\frac{1}{2} \sum_{i,j}
R_i(\theta^\alpha) [C_{ij}(\eta^A)]^{-1}
R_j(\theta^\alpha) \Bigr\}
\end{multline}
with respect to the timing-model parameters $\theta^\alpha$ \emph{and}, at the same time, the noise parameters $\eta^A$.

However, it is often desirable to include some \emph{a priori} notion of the nature and amplitude of the noise processes, and to weigh the relative plausibility of noise parameters over broad ranges instead of considering only their maximum-likelihood values---in other words, it is desirable to apply the framework of \emph{Bayesian inference} \citep{2010blda.book.....G} to the pulsar-timing problem, by evaluating the joint posterior probability
\begin{equation}
p(\theta^\alpha,\eta^A|t^\mathrm{obs}_i) =
\frac{p(t^\mathrm{obs}_i|\theta^\alpha,\eta^A) \, p(\theta^\alpha) \, p(\eta^A)
}{
\int p(t^\mathrm{obs}_i|\theta^\alpha,\eta^A) \, p(\theta^\alpha) \, p(\eta^A) \,
\mathrm{d} \theta^\alpha \, \mathrm{d} \eta^A
},
\end{equation}
where $p(\theta^\alpha)$ and $p(\eta^A)$ are \emph{prior probabilities} for the timing-model and noise parameters, respectively \citep{2009MNRAS.395.1005V}, and where the denominator is the Bayesian \emph{marginal likelihood} (or \emph{evidence} \citep{2010blda.book.....G}) for the \emph{type} of timing model that we are using, such as one that includes a certain parametrization of binary dynamics.

\paragraph*{Bayesian inference: motivation.} In this paper, we advocate the Bayesian approach even in cases where the noise parameters are assumed to be known in advance, because it allows us to include prior information about the timing-model parameters, and to obtain reliable parameter uncertainties (as the moments or quantiles of the posterior distribution\footnote{The least-squares errors can be construed as either \emph{frequentist} errors that describe the distribution of the maximum-likelihood estimator across noise realizations, or as \emph{Bayesian} uncertainties that describe the shape of the posterior probability when priors are negligible. Both kinds of errors are reliable only if the model of the data is linear in its parameters, or if its linearization is accurate throughout the region spanned by the predicted errors \citep{2008PhRvD..77d2001V}. Full Bayesian inference will of course provide exact Bayesian uncertainties; see \citet{2011PhRvL.107s1104V} for an approach to deriving exact frequentist errors.}) even when the linearized-residual approximation that underlies Eq.\ \eqref{eq:leastsquares} is not valid throughout the region of parameter space where the likelihood has its support.
While common lore is that this should never be the case in high-SNR observations such as pulsar timing, it may yet happen when the timing model includes parameters [or combinations of parameters, as witnessed by their very high correlation \citep{2006MNRAS.369..655H}] that have very weak but non-negligible effects on the residuals; the corresponding uncertainties can be so large that they take the parameters outside the ranges where their effects on the residuals can be approximated by the first derivative alone, or even outside their physical ranges.

As argued by one of us \citep{2008PhRvD..77d2001V}, the consequence is not just that the fit includes a bad parameter that should be ignored: the badness can be contagious by way of parameter correlations and affect even well-determined parameters with strong effects on the residuals.
In short, the worry is that the many digits of accuracy claimed for certain parameters in some pulsar ephemerides may be only apparent; in practice, however, this seems to be the exception.
% Reference \citep{2008PhRvD..77d2001V}
\citet{2008PhRvD..77d2001V} proposes a test (formulated in the language of Fisher matrices and GW observations, but applicable to this context) to verify empirically that the least-squares parameter uncertainties for an observation are compatible with the high-SNR, linearized-signal regime that they assume. If that is not so, fully nonlinear Bayesian inference is the only cure.

As we mentioned in the introduction, there are two more applications where a Bayesian approach can be useful: choosing between competing timing-model \emph{types} by way of Bayesian model comparison \citep{2010blda.book.....G}, which relies on evaluating marginal likelihoods; and validating the linearized-residual approximation used in pulsar-noise-- and GW--estimation methods that marginalize the posterior over the timing-model parameters, which we discuss next.

\paragraph*{Timing-model marginalization.}
Under the assumption that the $\theta^\alpha$ posteriors obtained in Bayesian inference are compatible with linearizing the residuals around a fiducial point (such as the white-noise maximum-likelihood estimator $\hat{\theta}^\alpha$), Eq.\ \eqref{eq:likelihood} can be rewritten as
\begin{multline}
\label{eq:linlikelihood}
p(t^\mathrm{obs}_i|\delta \hat{\theta}^\alpha + \delta \theta^\alpha,\eta^A) = \frac{1}{\sqrt{(2 \pi)^n |C(\eta^A)|}} \\
\times \exp \Bigl\{-\frac{1}{2}
[\hat{R}_i + \hat{M}_{i\beta} \delta \theta^\beta]
[C_{ij}(\eta^A)]^{-1}
[\hat{R}_j + \hat{M}_{j\gamma} \delta \theta^\gamma]\Bigr\}
\end{multline}
where to save space we denote $\hat{R}_i = R_i(\hat{\theta}^\alpha)$ and $\hat{M}_{i\beta} = M_{i\beta}(\hat{\theta}^\alpha)$. R.\ van Haasteren and colleagues \citep{2009MNRAS.395.1005V} realized\footnote{Or rather, they rediscovered a well-known result in the theory of Gaussian processes for machine learning \citep[sec.\ 2.7]{Rasmussen2005}.} that it is possible to integrate Eq.\ \eqref{eq:linlikelihood} analytically with respect to $\delta \theta^\alpha$, yielding
\begin{multline}
\label{eq:intlikelihood}
\int p(t^\mathrm{obs}_i|\delta \hat{\theta}^\alpha + \delta \theta^\alpha,\eta^A) \, \mathrm{d} (\delta \theta^\alpha) = \\
= \frac{
\exp \Bigl\{-\frac{1}{2}
\hat{R}_i
C'_{ij}(\eta^A)
\hat{R}_j \Bigr\}
}{\sqrt{(2 \pi)^{n - m} |C(\eta^A)| |\hat{M}^T C^{-1}(\eta^A) \hat{M}|}}
\end{multline}
where $C' = C^{-1} - C^{-1} \hat{M} (\hat{M}^T C^{-1} \hat{M})^{-1} \hat{M}^T C^{-1}$. If the prior $p(\theta^\alpha)$ is broad enough that it can be considered constant across the range of $\theta^\alpha$ that supports most of the likelihood, Eq.\ \eqref{eq:intlikelihood} can be used as the (quasi)likelihood in the expression for the marginalized $\eta^A$ posterior,
\begin{equation}
\label{eq:marlikelihood}
p(\eta^A|t^\mathrm{obs}_i) = \frac{\big[\int p(t^\mathrm{obs}_i|\delta \theta^\alpha,\eta^A) \, \mathrm{d}(\delta \theta^\alpha)] p(\eta^A)}{
\int \big[\int p(t^\mathrm{obs}_i|\delta \theta^\alpha,\eta^A) \, \mathrm{d}(\delta \theta^\alpha)\big] p(\eta^A) \, \mathrm{d}\eta^A}.
\end{equation}
This equation is clearly expedient if the goal is to estimate the pulsar-noise (or GW) parameters $\eta^A$ rather than the $\theta^\alpha$, since it replaces an integral over a moderately large number of dimensions with one over a much smaller-dimensional space.

For our purpose of seeking posterior distributions for the $\theta^\alpha$ subject to nontrivial priors, we note that it is possible to compute the integral of Eq.\ \eqref{eq:intlikelihood} over an $m_1$-dimensional subset $\theta^{\alpha_1}$ of ``boring'' timing-model parameters (for which we trust that the linearized-parameter approximation is accurate), obtaining a marginalized posterior that is a function of $\eta^A$ and of the remaining ``interesting'' timing-model parameters $\theta^{\alpha_2}$. For that, we simply replace $\hat{M}$ with the smaller matrix $M_1([\hat{\theta}^{\alpha_1},\theta^{\alpha_2}])$ obtained by taking only the columns of $M$ that correspond to the $\theta^{\alpha_1}$. 

\section{Computational setup}
\label{sec:compute}

We perform our tests of Bayesian inference and compare them with linearized least-squares results using the following computational setup.
\paragraph*{Likelihood.} We use G.\ Hobbs and R.\ Edwards' software package \textsc{tempo2} \citep{tempo2web,2006MNRAS.369..655H,2006MNRAS.372.1549E,2009MNRAS.394.1945H} to compute residuals and design matrices as a function of pulsar parameters. Whereas the standard mode of operation for \textsc{tempo2} is to compute the residuals and their derivatives once, and then perform least-squares fitting, we instead call \textsc{tempo2} repeatedly over many parameter sets to evaluate exact likelihoods across parameter space. To do so, we employ the Python wrapper \textsc{libstempo} \citep{libstempo}, written by one of us, which uses \textsc{tempo2} as a library, avoiding the overhead of repeated initialization for the same dataset.
\paragraph*{Sampling.} Using the residuals and the nominal TOA uncertainties (multiplied by the noise parameter EFAC), we can evaluate the likelihood of the data. We let EFAC vary as one of our search parameters, but we do not include any other noise parameters. We sample the posterior parameter distributions using the nested-sampling \citep{2004AIPC..735..395S} integrator \textsc{MultiNest} \citep{2009MNRAS.398.1601F}, which can be run in parallel over a moderate number of CPUs, and which returns an ``equal-weight'' sampling of the posteriors---that is, a population of pulsar parameter sets that can be immediately histogrammed to display marginalized distributions and averaged to provide conditional-mean parameter estimates.

The primary \textsc{MultiNest} product is actually the integrated evidence as a function of the data, likelihood, and of a set a parameter priors. The technique of nested sampling consists of evolving a cloud of ``live points'' in parameter space so that it climbs toward higher and higher likelihoods. The corresponding change in the ``prior mass'' at a certain likelihood value can be characterized statistically, yielding a Monte Carlo estimate of the evidence integral. We access \textsc{MultiNest} using the Python wrapper \textsc{PyMultiNest} \citep{PyMultiNest}.
\paragraph*{Driver.} Our code \textsc{bayesfit} \citep{bayesfit}, a single Python command-line application, pulls together the \textsc{tempo2} and \textsc{MultiNest} components, providing additional functionality such as the specification of priors; the Nelder--Mead optimization \citep{Nelder01011965}, \emph{post} sampling, of the maximum-posterior point in the final nested-sampling cloud; and the capability of computing the partially marginalized likelihood [Eq.\ \eqref{eq:intlikelihood}] for a given subset of timing-model parameters. Indeed, we find it convenient to integrate analytically over the \textsc{tempo2} dispersion-measure corrections and multi-frequency jump parameters. The former fit for the time dependence of dispersion (the $1/f^2$ delay incurred by pulse components of different frequencies as they travel through the interstellar medium); the latter fit for misalignments between the pulse-profile templates used to build TOAs from observations at different frequencies.
\paragraph*{Performance.} On our workstation (a dual quad-core 2.93 GHz Intel Xeon Mac Pro), a single evaluation of the likelihood, including the construction of residuals, takes 6 ms on a single core for a dataset of $\sim$ 200 TOAs. Adding partial analytical marginalization, including the required recomputation of the design matrix at each point, results in a single-likelihood cost of 60 ms, again for a dataset of $\sim$ 200 TOAs. A full \textsc{MultiNest} run routinely samples tens of thousands of parameter sets, yielding total execution times on the order of minutes to hours.
% B1953 parameter set: RAJ,DECJ,F0,F1,PMRA,PMDEC,PX,PB,T0,A1,OM,ECC
%
\paragraph*{\textsc{tempo2}.} To provide a comparison with linear least squares, we also run \textsc{tempo2} in its default mode, which weights the TOAs by the nominal errors. This produces best-fit estimates of timing-model parameters with uncertainties that are divided by the square root of the reduced $\chi^2$. It also produces the noise-weighted rms residual $\bar{R}$, which is usually slightly \emph{smaller} than the same quantity for the best-fit \textsc{bayesfit} solution. The reason is that the smallest $\bar{R}$ corresponds to the maximum-likelihood, but not necessarily maximum-posterior solution; furthermore, \textsc{tempo2} computes $\bar{R}$ using linear--least-squares updates of the input residuals, whereas \textsc{bayesfit} recomputes the residuals at the maximum-posterior location, which can lead to small numerical differences.

\section{Three case studies of Bayesian model fitting}
\label{sec:results}

\begin{table*}
\caption{Summary of Observing Properties and Timing Model Fits.}
\centering
\begin{threeparttable}
\begin{tabular}{| c | c | c c c | c c | c  c | c |}
	\hline
	Pulsar & Model & \multicolumn{3}{ |c| }{\# of parameters} & DM & DM Dist.\tnote{a} & \multicolumn{2}{ |c| }{Best-Fit Res.\ rms ($\mu\mathrm{s}$)} & Best-Fit $\chi^2$ \\
	 & & DM & Profile & Other\tnote{b}
 & ($\mathrm{pc} \: \mathrm{cm}^{-3}$) & (kpc) & \tempo & \bayesfit & \tempo \\
	\hline
	J1640+2224 & DD & 23 & 26 & 12 & 18.426 & 1.16 & 0.562 & 0.565 & 4.35 \\
	B1953+29 & DD & 0 & 27 & 12 & 104.5 & 4.64 & 3.980 & 4.015 & 0.98 \\
	J2317+1439 & ELL1 & 30 & 12  & 15 & 21.9 & 0.83 & 0.496 & 0.508 & 3.03 \\
	\hline
\end{tabular}
\begin{tablenotes}
	\item[a] DM distance is calculated using the NE2001 map of electron density \citep{2002astro.ph..7156C}.
	% \item[b] Datasets and \tempo~timing solutions come from the NANOGrav dataset analyzed in \citep{2013ApJ...762...94D}.
	\item[b] ``Other'' includes the spin, astrometric, and binary parameters.
\end{tablenotes}
\end{threeparttable}
\label{tab:psr_summary}
\end{table*}
\begin{table*}
\caption{Timing Solutions for J1640+2224, B1953+29, and J2317+1439 from \tempo.}
\centering
\begin{threeparttable}
\begin{tabular}{| l | c c c |}
	\hline
	Parameter & J1640+2224 & B1953+29 & J2317+1439 \\
	\hline
	Right ascension, $\alpha$ (J2000.0) & 16:40:16.74350(6) & 19:55:27.87596(3) & 23:17:09.23701(1) \\
	Declination, $\delta$ (J2000.0) & +22:24:08.9433(7) & +29:08:43.4656(5) & +14:39:31.2449(3) \\
	Spin frequency, $\nu$ ($\mathrm{s}^{-1}$) & 316.12398431362(3) & $163.047913069140(3)$ & $290.254608187112(3)$ \\
	Spin frequency derivative, $\dot{\nu}$ ($\mathrm{s}^{-2}$) & $-2.812(1)\times10^{-16}$ & $-7.907(3)\times10^{-16}$ & $-2.0475(6)\times10^{-16}$ \\
	Proper motion, $\mu_\alpha$ ($\mathrm{mas}\:\mathrm{yr}^{-1}$) & $2.2(1)$ & $-1.3(3)$ & $-0.8(2)$ \\
	Proper motion, $\mu_\delta$ ($\mathrm{mas}\:\mathrm{yr}^{-1}$) & $-11.0(1)$ & $-3.9(4)$ & 3.1(3) \\
	Parallax, $\pi$ (mas) & & $-3(2)$ & $-0.7(4)$ \\
	Orbital period, $P_b$ (days) & $175.46066225(7)$ & $117.3490971(1)$ & $2.4593314628(3)$ \\
	Orbital period derivative, $\dot{P_b}$ & & & $6.4(9)\times10^{-12}$ \\
	Epoch of periastron, $T_0$ (MJD) & $51626.1796(4)$ & $46112.477(3)$ & \\
	Epoch of ascending node, $T_\mathrm{asc}$ (MJD) & & & 54000.25476694(5) \\
	Projected semimajor axis, $x$ (lt-sec) & $55.3297183(7)$ & $31.4126902(6)$ & $2.3139480(3)$ \\
	Longitude of periastron, $\omega$ (deg) & $50.7330(9)$ & $29.474(7)$ & \\
	Eccentricity, $e$ & $7.9712(3)\times10^{-4}$ & $3.3016(4)\times10^{-4}$ & \\
	$\epsilon_1 \equiv e \sin\omega$ & & & $6(2)\times10^{-7}$ \\
	$\epsilon_2 \equiv e \cos\omega$ & & & $1(3)\times10^{-7}$ \\
	$\dot{\epsilon_1}$ & & & $-2(4)\times10^{-15}$ \\
	$\dot{\epsilon_2}$ & & & $2.0(7)\times10^{-14}$ \\
	Orbital inclination, $\sin i$ & $0.99$\tnote{a} & & \\
	Companion mass, $m_2$ ($M_\odot$) & $0.25(4)$ & & \\
	\hline
\end{tabular}
\begin{tablenotes}
	\item[a] Inclination angle is held fixed at $\sin i=0.99$.
\end{tablenotes}
\end{threeparttable}
\label{tab:timing}
\end{table*}

As a test of our method, we ran \bayesfit~on the pulsars included in the NANOGrav 2012 dataset \citep{2013ApJ...762...94D} and compared our timing solutions to the best-fit values, errors, and timing residuals obtained with linear least-squares fits (i.e., by \tempo). In most cases, the solutions were consistent, but in a few cases we found interesting discrepancies.
In this section, we describe the results of Bayesian inference for three pulsars (B1953+29, J2317+1439, and J1640+2224), which we selected to demonstrate different benefits of Bayesian inference.
In the case of B1953+29, we see that Bayesian analysis confirms and validates the least-squares result.
In the case of J2317+1439, we show that incorporating prior information about a poorly determined parameter (the parallax) fixes the overall parameter-estimation bias engendered by excluding it from the model, or by accepting its unphysical least-squares estimate.
In the case of J1640+2224, we show that a full Bayesian analysis produces a timing solution that is rather different from the the linear least-squares estimate, especially for the binary parameters. Suggestively, the posterior distribution for the mass $M_2$ of the pulsar companion, which has been identified optically as a white dwarf (\citealp*{1996ASPC..105..497L}; \citep{1996ApJ...458L..33L}), attributes $\sim 35\%$ probability to $M_2$ above the Chandrasekhar limit. Using the optical observations to place a prior on $M_2$ changes the picture considerably.

The TOAs used here are part of a set of NANOGrav observations made between 2005 and 2010 in a program to measure or constrain the GW stochastic background \citep{2013ApJ...762...94D}. The pulsars considered here were all observed using the 305-m NAIC Arecibo Observatory (AO). Table~\ref{tab:psr_summary} summarizes the timing results. All of the timing solutions include the position, proper motion, spin frequency, and spin frequency derivative. All three pulsars are in binary systems, so the timing models include the five Keplerian binary parameters [fit using either the ``DD'' \citep{1985AIHS...43..107D} or ``ELL1'' \citep{2001MNRAS.326..274L} timing models], plus additional post-Keplerian parameters if they improve the fit. In addition, the timing models include parameters describing time variations in the DM and frequency-dependent changes in the pulse-profile shape, which are implemented as jumps between frequencies.

For all three of these pulsars, either the \tempo~value of the parallax (PX) is consistent with zero, or the parallax is not included in the fit. Indeed, parallax is notoriously difficult to fit using pulsar timing.
With Bayesian inference, however, we can include a prior on the parallax. Since only one of the pulsars in the NANOGrav has a published parallax measurement from VLBI,\footnote{J1713+0747, from \cite{2009ApJ...698..250C}} we defined the prior on the parallax by estimating the distance from the pulsar's dispersion measure (DM). We used the ``NE2001'' map of the electron density in the Galaxy \citep{2002astro.ph..7156C}, and assumed that the NE2001 DM distance $d$ has a Gaussian distribution with variance $\sigma^2_d$. This corresponds to the prior
\begin{equation}
p(\mathrm{PX}) = \frac{1}{\sqrt{2\pi} \, \sigma_d \, \mathrm{PX}^2} \exp\left[ -\frac{(\mathrm{PX}^{-1}-d)^2}{2\sigma_d^2} \right], \quad \mathrm{PX} > 0. \label{eq:px_distribution}
\end{equation}
For simplicity, we set $\sigma_d = 0.2 d$ \citep{Lazio}.
% Additionally, J1640+2224 had other interesting properties, which we discuss in more detail below.

\subsection{PSR B1953+29}
\begin{figure*}
\begin{center}
\includegraphics[width=\textwidth]{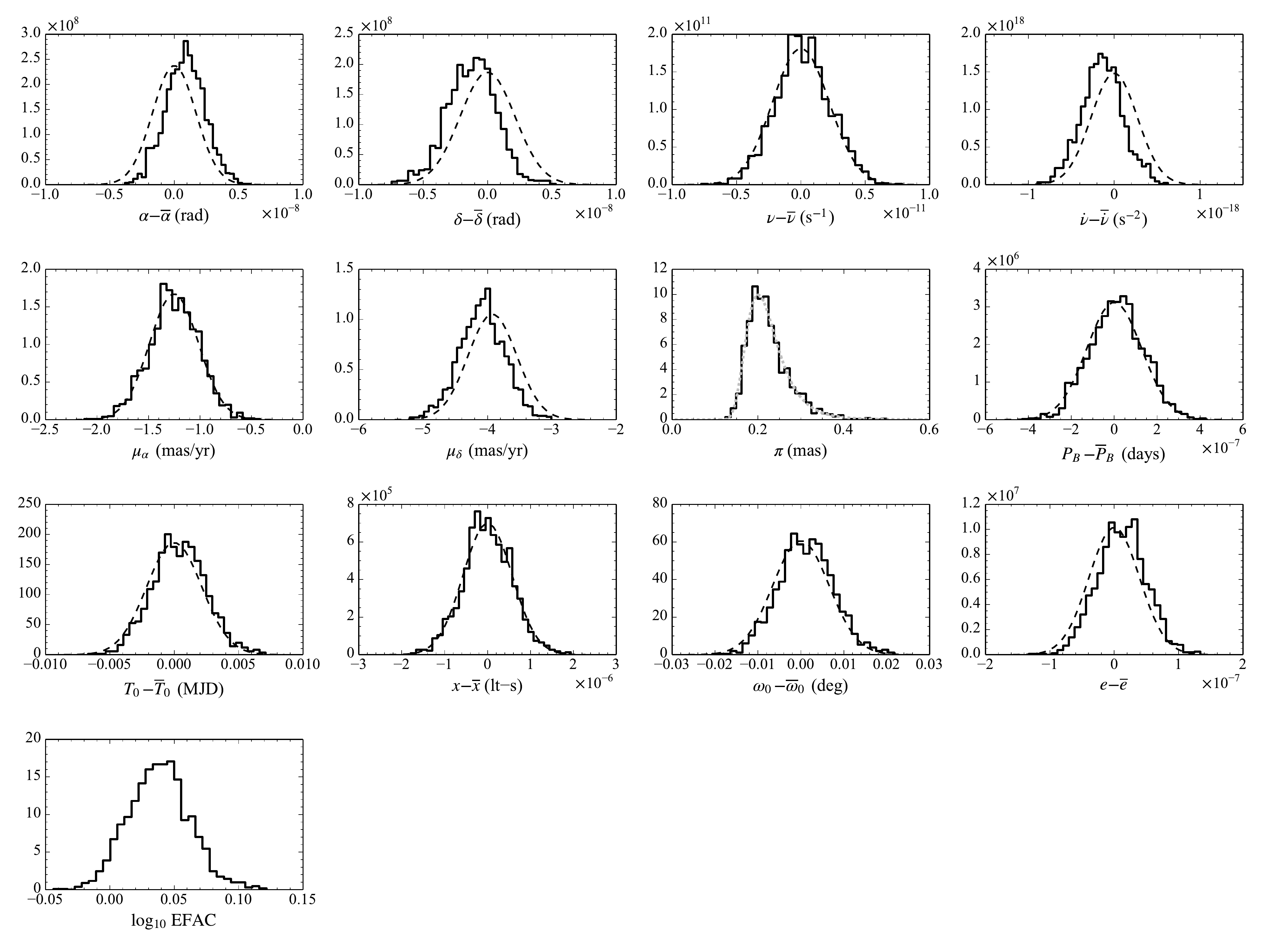}
\caption{Posterior probability distributions (solid step lines) of the timing parameters for PSR B1953+29. The TOAs were fit using the DD model. A bar over a parameter refers to the parameter value obtained using \tempo, which is listed in Table~\ref{tab:timing}. The dashed curves show the best-fit values and errors obtained using \tempo, and the light dotted curve in the parallax plot shows the prior probability distribution, which was derived from the DM distance. The posterior distribution for the parallax is the same as the prior, indicating that the timing observations do not provide any additional information. The posterior distributions for the other parameters agree with the \tempo~timing solution. The \tempo~solution for the parallax is $\pi=-2.7\pm 2.4$ mas; we do not plot the corresponding distribution because most of it the lies outside of the \bayesfit\ prior and posterior distributions.}
\label{fig:B1953+29}
\end{center}
\end{figure*}

The observed TOAs for B1953+29 are fit using the DD binary pulsar model with 12 spin, astrometric, and binary parameters. We also fit the parameter $\log_{10}\mathrm{EFAC}$, which rescales the TOA errors. Additionally, the timing solution includes 27 parameters describing pulse-profile variations: the linearized-residual approximation is very accurate for these, so we marginalize the posterior over them using Eq.\ \eqref{eq:intlikelihood}. As described in Eq.\ \eqref{eq:px_distribution}, the prior on the parallax is defined on the basis of the DM distance of $4.64\:\mathrm{kpc}$. For all of the other parameters, we assume a uniform prior centered around the best-fit value given by \tempo, with width chosen to include the entire posterior.

Figure~\ref{fig:B1953+29} shows the posterior probability distributions (the solid step lines) for the spin, astrometric, and binary parameters and for $\log_{10}\mathrm{EFAC}$, as well as the corresponding \tempo\ distributions (the dashed curves). The best-fit parameter values found with \tempo~are listed in Table~\ref{tab:timing}. The most significant difference in the \tempo~and \bayesfit\ solutions is in the value of the parallax. In the \tempo~timing solution, the parallax is given as $\pi = -2.7 \pm 2.4 \: \mathrm{mas}$, which is obviously unphysical. (In these units, the parallax is equal to the reciprocal of the conventional ``parallax distance,'' given in kiloparsecs.) In the \bayesfit\ solution, the posterior for the parallax is identical to the prior; in other words, the timing observations do not provide any additional information about the parallax. The posteriors for the other parameters agree well with the \tempo~values. The rms of the residuals is slightly larger, $4.015\:\mathrm{\mu s}$ with \bayesfit\ (using the posterior mode) versus $3.980 \: \mathrm{\mu s}$ with \tempo.

We conclude that in this case including a prior on the parallax does not change the timing solution. Using \bayesfit~allows us to validate the linear least-squares solution found with \tempo~and confirms that the parameters' posterior probability distributions have a Gaussian distribution.

\subsection{PSR J2317+1439}
\begin{figure*}
\begin{center}
\includegraphics[width=\textwidth]{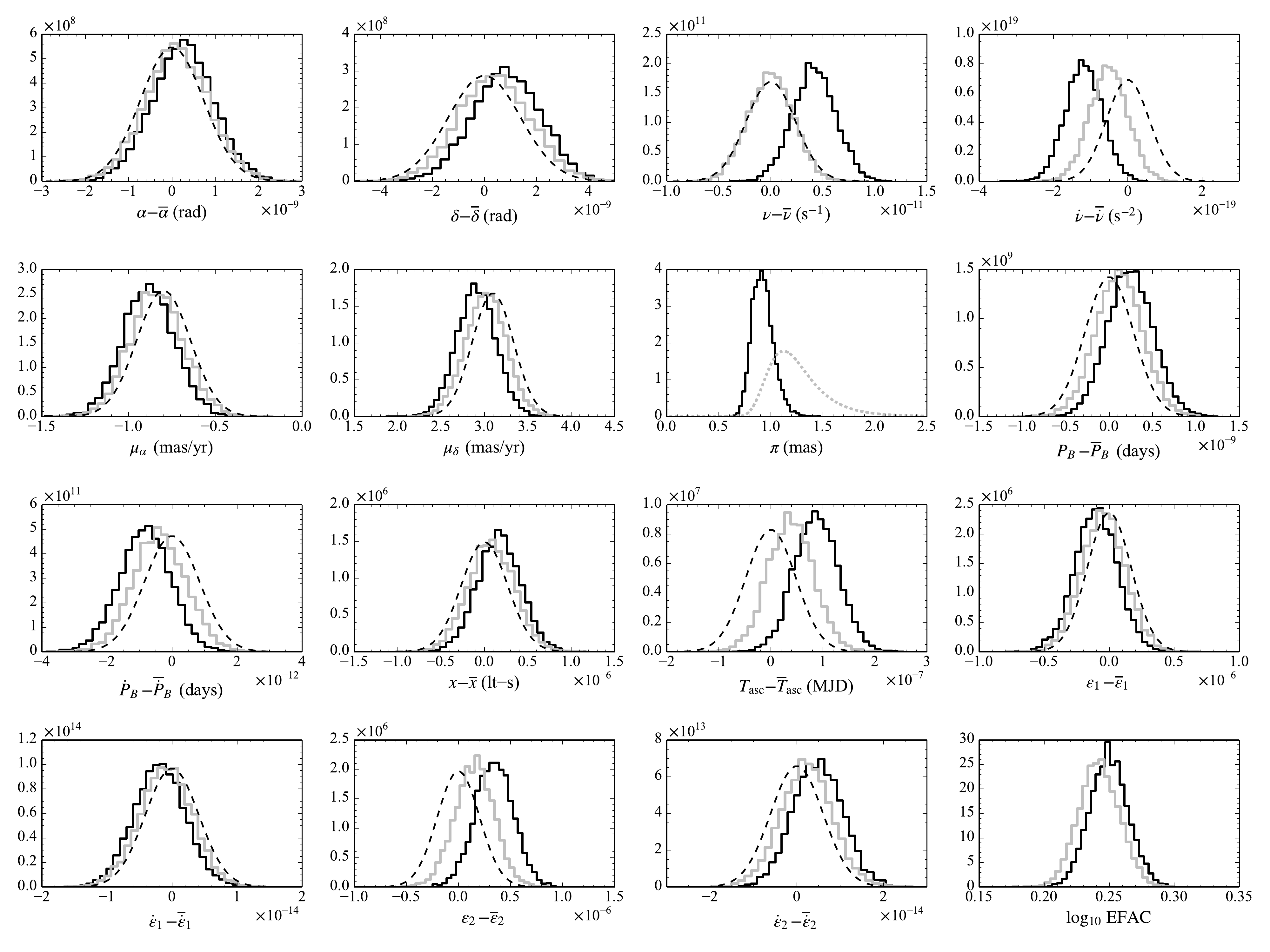}
\caption{Posterior probability distributions (solid step lines) of the timing parameters for PSR J2317+1439. The TOAs were fit using the ELL1 model. A bar over a parameter refers to the parameter value obtained using \tempo, which is listed in Table~\ref{tab:timing}. The darker step lines show the posteriors obtained by placing a DM-distance prior (the dotted curve) on $\pi$; the lighter step lines show the posteriors obtained by setting $\pi$ to zero. The dashed curves show the \tempo\ normal distributions with $\pi=-0.74 \pm 0.38$ (which here lies outside the $\pi$ plot). A \tempo\ solution with $\pi$ set to 0 would lie exactly along the corresponding Bayesian solution. The three timing solutions show significant differences in several parameters.}
\label{fig:J2317+1439}
\end{center}
\end{figure*}

The observed TOAs for J2317+1439 are fit using the ELL1 timing model with 15 spin, astrometric, and binary parameters. We also fit the parameter $\log_{10} \mathrm{EFAC}$, and marginalize posteriors analytically over 42 parameters describing DM variations and pulse-profile variations. The \tempo~timing solution gives a small (and unphysically negative) value of the parallax, $\pi = -0.74 \pm 0.38 \; \mathrm{mas}$; however, the low DM of $21.9 \; \mathrm{pc} \; \mathrm{cm}^{-3}$ implies that this pulsar has a DM distance of only $0.83 \: \mathrm{kpc}$ and a parallax of $1.2 \; \mathrm{mas}$. To explore the timing-model effects of the parallax, we run two variants of our Bayesian analysis: one in which we impose a DM-distance prior on $\pi$, and another in which we omit $\pi$ from the timing model (which is equivalent to setting it to zero). For all of the other parameters, we assume a uniform prior centered around the best-fit value given by \tempo, with width chosen to include the entire posterior.

Figure~\ref{fig:J2317+1439} compares the \bayesfit~posterior probability distributions (where the darker solid lines correspond to the DM-distance prior on $\pi$, and the lighter solid lines to setting $\pi = 0$) with the best-fit normal distributions found with \tempo\ (which include $\pi = -0.74 \pm 0.38$, and are shown as dashed lines).
In the prior-on-$\pi$ analysis, the combination of the prior and of the timing data produces a posterior that favors smaller values of the parallax than the prior. A number of the other parameters are shifted with respect to the \tempo\ analysis, particularly the spin frequency and spin frequency derivative. The timing residuals are only slightly different for the two solutions, with an rms of $0.500\:\mu\mathrm{s}$ for the \bayesfit\ posterior-mode solution and $0.496\:\mu\mathrm{s}$ for the \tempo\ best-fit solution. The posterior-mode value of the EFAC parameter is $1.77$, which is in good agreement with the reduced $\chi^2$ value from the \tempo\ analysis ($\mathrm{EFAC}^2 = 3.12$, compared with $\chi^2 = 3.03$).
In the setting-$\pi$-to-zero analysis, the timing parameters are again shifted; in fact, their distributions are entirely consistent with a \tempo\ run where $\pi$ is set to zero (not shown here).

Thus, we see that either accepting an unphysical best-fit value for $\pi$ or excluding it from the timing model leads to a slight but non-negligible bias in a number of other parameters; the bias is corrected in the Bayesian-inference analysis by imposing a physically motivated prior on the parallax. Indeed, in this case it would be useful to measure the parallax more reliably and accurately using VLBI.

\subsection{PSR J1640+2224}
\begin{figure*}
\begin{center}
\includegraphics[width=\textwidth]{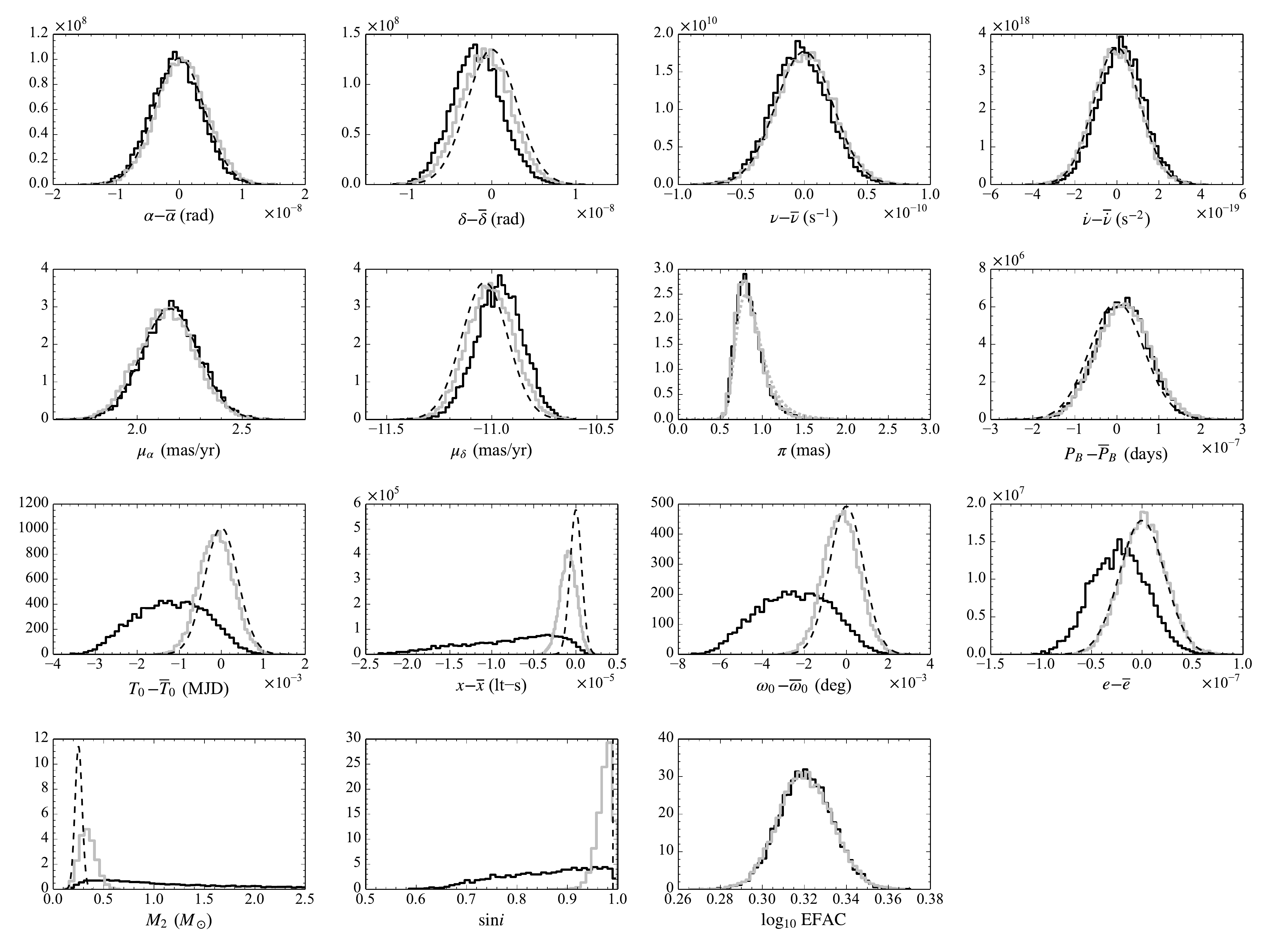}
\caption{Posterior probability distributions (solid step lines) of the timing parameters for PSR J1640+2224. The TOAs were fit using the DD model. A bar over a parameter refers to the parameter value obtained using \tempo, which is listed in Table~\ref{tab:timing}. The darker step lines show the posteriors obtained by placing a uniform prior on $M_2$; the lighter step lines show the posteriors obtained by adopting a ``white-dwarf'' prior for $M_2$ (a normal distribution centered on $0.3 \: M_\odot$ with $0.1 \: M_\odot$ standard deviation). The dashed curves show the \tempo\ normal distributions (and the dashed vertical line shows the fixed \tempo\ value for $\sin i$), while the dotted curve traces the DM-distance prior for the parallax. There are significant differences between the three timing solutions in the values of the binary parameters.\label{fig:J1640+2224}}
\end{center}
\end{figure*}

J1640+2224 deserves special consideration as one of a few systems where it was possible to measure the Shapiro delay \citep{2005ApJ...621..388L}. L\"ohmer and colleagues report a companion mass $M_2 = 0.15^{+0.08}_{-0.05} \, M_\odot$ and a Shapiro-delay \emph{shape} $\sin i \gtrsim 0.98$ (where $i$ is the orbital inclination).
Furthermore, % Lundgren, Foster, and Camilo \citep{1996ASPC..105..497L}
\citet{1996ASPC..105..497L,1996ApJ...458L..33L} identify an optical counterpart to J1640+2224 with properties consistent with a white dwarf of mass $\simeq 0.3 \: M_\odot$.

The DD timing solution for J1640+2224 adopted by \citet{2013ApJ...762...94D} fits for 13 parameters describing the spin, astrometric, and binary properties of the system (including $M_2$ but not $\sin i$), and for 49 parameters describing the DM variations and pulse-profile variations with frequency.
The parameter $\sin i$ cannot be fit by linear least squares, which would report an unphysical value of $i$ for which the timing model cannot be evaluated, so Demorest and colleagues fix $\sin i$ to 0.99, which yields a best-fit $M_2 \simeq 0.3 \: M_\odot$.
In our analysis we fit for the 14 spin, astrometric, and binary parameters (including both $M_2$ and $\sin i$) plus the parallax, which is not included in the \citet{2013ApJ...762...94D} timing solution.

To explore the timing-model effects of incorporating information from optical observations, we run two variants of our Bayesian analysis, one in which we impose physical constraints on $i$ and $M_2$ (with uniform priors $\cos i \in [0,1]$ and $M_2 \in [0,10] \, M_\odot$), and another in which we further constrain $M_2$ with a ``white-dwarf'' prior (a normal distribution centered on $0.3 \: M_\odot$ with standard deviation $0.1 \: M_\odot$). As in the previous examples, we integrate over the parameters describing the DM variations and pulse-profile shape, and we use a prior distribution for the parallax derived from the dispersion measure, which corresponds to a DM distance of $1.16 \; \mathrm{kpc}$ and a parallax of $0.63 \; \mathrm{mas}$.
For all of the other parameters, we assume a uniform prior centered around the \tempo~best-fit values with a sufficiently wide range to encompass the entire posterior.

Figure~\ref{fig:J1640+2224} compares the posterior probability distributions calculated with \bayesfit\ (the darker solid step lines for the uniform-$M_2$ run, and the lighter set for the white-dwarf--prior run) and the \tempo\ normal distributions (the dashed curves). 
The posteriors for the parallax are very close to the prior (the dotted curve). 
There is significant disagreement between the \bayesfit\ uniform-$M_2$ and the \tempo\ timing solutions in the values of the binary parameters, some of which have non-Gaussian distributions. By contrast, the \bayesfit\ white-dwarf--prior distributions are closer to the \tempo\ solutions. The residuals however are very similar: for the uniform-$M_2$ solution, the mode of the \bayesfit\ posterior corresponds to residual rms of $0.565\:\mu\mathrm{s}$ compared to the $0.562\:\mu\mathrm{s}$ obtained with the \tempo~solution, and the posterior for EFAC is in good agreement with the best-fit \tempo\ value of reduced $\chi^2$ ($\mathrm{EFAC}^2=4.47$ vs.\ $\chi^2 = 4.35$).

\begin{figure}
\begin{center}
\includegraphics[width=0.65\columnwidth]{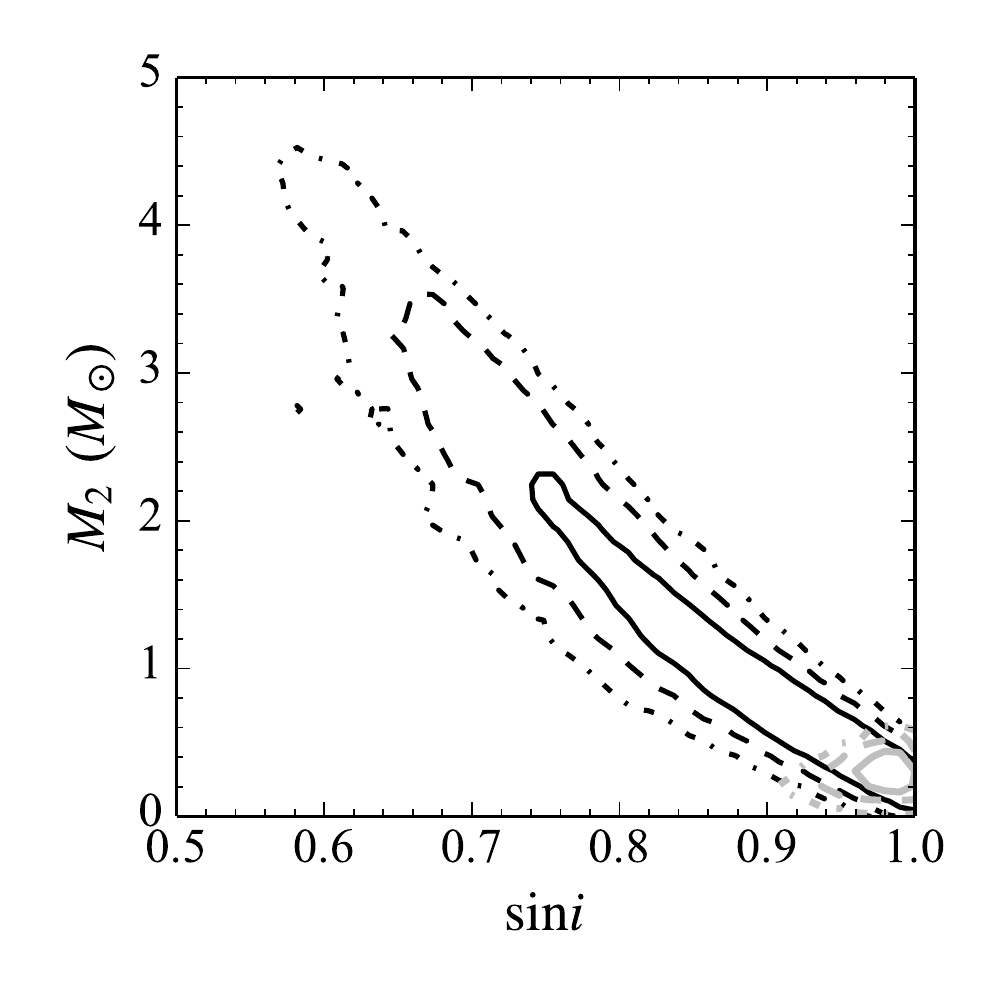}
\end{center}
\caption{Joint posterior probability distributions of the timing parameters $M_2$ and $\sin i$ for PSR J1640+2224. The darker and lighter lines show the posteriors obtained with uniform and ``white-dwarf'' $M_2$ priors, respectively. The solid, dashed, and dashed-dotted lines trace the 68\%, 95\%, and 99\% confidence intervals.\label{fig:J1640+2224_2Dhist}}
\end{figure}

In the uniform-$M_2$ solution, the $\sin i$ posterior prefers smaller values than reported by \citet{2005ApJ...621..388L}, corresponding to rather larger companion masses $M_2$. Indeed, the cumulative probability distribution for $M_2$ places only 5\% probability below $0.3 \: M_\odot$, and
35\% probability above the Chandrasekhar limit of $1.44 \: M_\odot$, so the timing data \emph{alone} is not fully compatible with the identification of the counterpart as a low-mass white dwarf. We plot the joint posterior distributions of $M_2$ and $\sin i$ in Fig.~\ref{fig:J1640+2224_2Dhist} (where again the darker lines correspond to the uniform-$M_2$ solution, and the lighter lines to the white-dwarf--prior solution). Clearly $M_2$ and $\sin i$ are poorly determined and rather correlated, which is not surprising since the measurement of both comes from the Shapiro delay, a weak effect in the timing solution. The white-dwarf--prior distribution is much more constrained, as expected.

This case demonstrates that a nonlinear Bayesian analysis can properly characterize timing solutions where a few parameters have weak but non-negligible effects, so both their nonlinear behavior and their acceptable physical ranges come into play, and a linear least-squares approach is insufficient. This case demonstrates also the potential impact of parameter priors from non-timing observations, which are accounted properly in the Bayesian framework. In the light of our reanalysis (especially when we allow for intrinsic pulsar noise, as we do in a forthcoming paper), it would be interesting to revisit the question whether the identification of the J1640+2224 companion is truly definitive.

\section{Conclusions}
\label{sec:conclusions}

In this paper we have advocated the use of fully nonlinear Bayesian inference to derive pulsar-timing solutions, and we have shown that doing so is already practical with current hardware and software. To wit, we have used Python wrappers [\textsc{PyMultiNest} by Johannes Buchner \citep{PyMultiNest} and \textsc{libstempo} by one of us \citep{libstempo}] to connect a powerful stochastic sampler [\textsc{MultiNest} \citep{2009MNRAS.398.1601F}] with a state-of-the-art pulsar-timing application [\textsc{tempo2} \citep{tempo2web}]. Our code, \textsc{bayesfit}, is available on GitHub for inspection, inspiration, and reuse \citep{bayesfit}.

In Sec.\ \ref{sec:timing} we have surveyed the mathematical formulation of timing-model fitting, and in Sec.\ \ref{sec:results} we have shown three practical examples where the nonlinear Bayesian analysis provides different benefits: for PSR B1953+29, it validates the linearized least-squares estimates; for PSR J2317+1439, it corrects for parameter-estimation bias by incorporating prior information about the pulsar distance; for PSR J1640+2224, it provides the proper characterization of non-Gaussian posterior parameter distributions, and it accounts for a strong parameter prior from optical observations.

In Sec.\ \ref{sec:intro} we have suggested two additional applications of Bayesian inference, which may be even more compelling, but which we did not explore practically in this paper.
%
%\begin{itemize}
%
% \item

First, the Bayesian framework is the appropriate formulation for simultaneously inferring the pulsar parameters that have deterministic effects on the timing model \emph{and} the statistical properties of noise from the pulsar and the detector. Indeed, it is not possible to study noise without simultaneously revisiting the timing model, a point made forcefully
% by van Haasteren and Levin \citep{2013MNRAS.428.1147V}.
by \citet{2013MNRAS.428.1147V}.
Furthermore, unmodeled noise and conceivably GWs will certainly skew the estimates of timing-model uncertainties and even the best-fit parameters \citep{temponest}.
%
%\item

Second, Bayesian model comparison can provide a quantitative way to choose between alternative timing models, such as one that includes modified-gravity corrections [or even planets \citep{1992Natur.355..145W}] vs.\ one that does not. Such a comparison is most significant when there is a physically well-defined way to attribute prior probabilities to the competing hypotheses, but in the absence of that it is possible to perform a frequentist analysis of a \emph{detection scheme} (i.e., detection of modified gravity, or planets) based on Bayesian evidence \citep{2012PhRvD..86h2001V}.

\citet{temponest} compute the evidence for different noise models and subsets of timing-model parameters, and use its relative values to select a preferred model. The way that they do so raises difficult issues that are typical of Bayesian inference problems for weak signals immersed in noise. For instance, some of the \citet{temponest} noise models include an ``optimal'' set of lines selected with an iterative procedure. The line-frequency priors should then (but do not) reflect all other sets that could have resulted from the process, and thus penalize the optimal set, a very fine-tuned choice that will perforce fit (or overfit) the data.

% In our judgment, however, they do so in a way that is not entirely Bayesian, or statistically principled. This occurs in their comparisons of both noise models and parameter sets. In the first case, we submit that the prior probabilities for the noise models that include lines are not well defined, because of the iterative procedure used to select the set of ``optimal'' lines: the line-frequency prior should reflect all the other possible choices of lines that could have resulted from such a process, and thus penalize the optimal set, a very fine-tuned choice that will perforce fit (or overfit) the data.

Furthermore, the procedure used by \citet{temponest}, whereby additional timing parameters are added to the model and retained only if the evidence improves, seems ill-suited to \emph{physical} parameters such as the first derivatives of the eccentricity and orbital periods. Doing so amounts to comparing models in which some of the physical consequences of general relativity are ``turned off.''
Omitting parameters from a model (which in \tempo\ equates to fixing them to zero) should be a question of \emph{approximation}, not model comparison: it should be done only when it is determined that the corresponding timing-model terms are negligibly small.\footnote{The same reasoning applies to the phase jumps applied between timings taken at different observatories. These jumps are always present, but could be deemed to be negligible on the basis of experimental considerations.}
Even if we accepted that such model comparisons are meaningful, they remain ill-defined in the absence of strongly motivated priors for the extra parameters. These priors affect the Occam-factor penalty incurred by the higher-dimensional models,\footnote{This is unless the extra parameters are determined so poorly that they populate the priors uniformly, in which case the evidence is unchanged.} so they can make the difference between winning or losing a comparison.
%
%\end{itemize}
%

We plan to explore these issues in future work.

\paragraph*{Acknowledgments.}
We are grateful to Paul Demorest, Joe Lazio, Sarah Burke--Spolaor, Rutger van Haasteren, Lindley Lentati, Tom Prince, and to all NANOGrav colleagues for helpful discussions. SV was supported by an appointment to the NASA Postdoctoral Program at the Jet Propulsion Laboratory administered by Oak Ridge Associated Universities through a contract with NASA. MV was supported by the Jet Propulsion Laboratory RTD program. This work was carried out at the Jet Propulsion Laboratory, California Institute of Technology, under contract to the National Aeronautics and Space Administration. Copyright 2014 California Institute of Technology.

\bibliography{phyjabb,master}

\end{document}